\documentclass[a4paper,10pt]{article}
\usepackage{amssymb,amsfonts,amsmath,amsthm}
\usepackage[margin=0.9in]{geometry}

\def\upd{\,{\rm d}}
\newcommand{\be}{\begin{eqnarray}}
\newcommand{\ee}{\end{eqnarray}}
\newcommand{\nnee}{\nonumber\ee}
\newcommand{\Tr}{\,{\rm Tr}\,}

\def\Co{{\mathbb C}}
\def\Io{{\mathbb I}}
\def\Mo{{\mathbb M}}
\def\Ro{{\mathbb R}}
\def\To{{\mathbb T}}

\newtheorem{Theorem}{Theorem}
\newtheorem{Proposition}{Proposition}
\newtheorem{Lemma}{Lemma}

\def\beginproof{\par\strut\vskip 0.0cm\noindent{\bf Proof}\par}
\def\endproof{\par\strut\hfill$\square$\par\vskip 0.2cm}

\title{Quantum Statistical Manifolds}

\author{
Jan Naudts\\
\strut\\
\small          Departement Fysica, Universiteit Antwerpen,\\
\small          Universiteitsplein 1, 2610 Antwerpen, Belgium\\
\small          ~~E-mail: jan.naudts@uantwerpen.be.
}
\date {}

\begin{document}
\maketitle

\abstract{
Quantum information geometry studies families of quantum states
by means of differential geometry. 
A new approach is followed with the intention to facilitate
the introduction of a more general theory in subsequent work.
To this purpose, the emphasis is shifted from a
manifold of strictly positive density matrices
to a manifold of faithful quantum states on the C*-algebra
of bounded linear operators. In addition, ideas from
the parameter-free approach to information geometry are adopted.
The underlying Hilbert space is assumed to be finite-dimensional.
In this way technicalities are avoided so that strong results are obtained,
which one can hope to prove later on in a more general context.
Two different atlases are introduced, one in which it is straightforward
to show that the quantum states form a Banach manifold, the other
which is compatible with the inner product of Bogoliubov
and which yields affine coordinates for the exponential connection.
}


\section{Introduction}

The basic example of a quantum statistical system starts with a self-adjoint operator $H$
on a finite-dimensional or separable Hilbert space $\cal H$, with the property that
the operator $\exp(-\beta H)$ is trace-class for all $\beta$ in an open interval $D$
of the real line $\Ro$. 
Then for $\beta\in D$ the quantum expectation value $\langle A\rangle_\beta$
of any bounded operator $A$ in $\cal H$ is given by
\be
\langle A\rangle_\beta&=&\frac{1}{Z(\beta)}\Tr e^{-\beta H} A,
\quad \mbox{ with }Z(\beta)=\Tr e^{-\beta H}.
\label{intro:bg}
\ee
Note that the quantum expectation is well-defined because the product of a trace-class operator
with a bounded operator is again a trace-class operator.
The operator $H$ is the {\em Hamiltonian}. It defines a one parameter family of quantum states
via (\ref{intro:bg}).

The quantum state (\ref {intro:bg}) is a simple example of a model
belonging to the {\em quantum exponential family}. In this case, the quantum states
form a one dimensional manifold.
The goal of the present work is to search for a quantum exponential family
formulated in a parameter-free way, similar to the formulation of Pistone and Sempi \cite{PS95}
in the non-quantum case, and to investigate a further generalization involving deformed
exponential functions along the lines set out by the author in \cite{NJ04}.

Early efforts to use geometric methods in the study of non-commut\-at\-ive information theory include the 
work by Ingarden and coworkers. See for instance \cite{IRS81,IJKK82}.
The relation with Amari's Information Geometry \cite{AS85,AN00} was studied by Hasegawa \cite{HH93, HH97}.
He introduced an alpha-family of divergences $D_\alpha(\rho,\sigma)$ where $\rho$ and $\sigma$
are any pair of density operators on a finite-dimensional Hilbert space $\cal H$.
The approach relies strongly on the properties of the trace.

The metric on the manifold of density matrices is the scalar product introduced by Bogoliubov
and used by Kubo 
and Mori 
in the context of linear response theory. The generalization to
an inner product for vector states on a von Neumann algebra is given in \cite{NVW75}.
See also \cite{PT93}.

A more recent account on quantum information geometry is found in Chapter 7 of \cite{AN00}.
See also Example 3.8 of \cite{PD08}.

The parameter-free approach of Pistone and Sempi \cite{PS95} was generalized to the quantum
context by Streater \cite{GS97,SRF04a,SRF04b}. Both the classical case
and the quantum case need a regularizing condition on the allowed density functions,
respectively density operators. Under this condition they form a Banach manifold.
Recently, Newton \cite{NN12,MP17} proposed an alternative regularization based
on a specific choice of a deformed logarithmic function. Part of the
arguments in \cite{MP17} can be transposed to the quantum setting \cite{NJ17}.

The structure of the paper is as follows.
In the next section quantum states are labeled with operators belonging to the commutant
of the GNS-representation rather than with density matrices. Section \ref {sect:tangent}
describes the plane tangent at the reference state. Next, an atlas is
introduced which contains a multitude of charts, one for each element of the manifold.
Theorem \ref {atlas:theorem} proves that the manifold is a Banach manifold and that
the cross-over maps are linear operators. Section \ref {sect:bogo} introduces
the inner product of Bogoliubov. The metric tensor is calculated.
Next alternative charts are introduced and their relation with the metric tensor is investigated.
Sections \ref {sect:mix} and \ref {sect:exp} discuss the mixture and the exponential connections.
Proposition \ref {exp:prop3} proves that the alternative charts provide affine coordinates for the
exponential connection. Section \ref {sect:modul} contains a short presentation of the
additional structure provided in quantum information geometry by the existence
of modular automorphism groups.
The final section discusses the results obtained so far.
An appendix about the GNS-representation and the modular operator is added for convenience
of the reader.

\section{Representation theorems}

In the present paper the Hilbert space $\cal H$ is assumed to be finite dimensional.
This solves the question of choosing an appropriate topology on the manifold of quantum states.
In addition, all operators under consideration are bounded continuous.
In fact they are finite-dimensional matrices. In this way the
technical difficulties of working with unbounded operators are avoided.

A density matrix $\rho$ is a self-adjoint operator with discrete spectrum
consisting of non-negative eigenvalues which add up to one. This implies that the 
trace satisfies $\Tr \rho=1$.
The operator $e^{-\beta H}/Z(\beta)$, mentioned in the Introduction is
a density matrix of the kind we have in mind.

Introduce the notation ${\cal A}={\cal B}({\cal H})$ for the $C^*$-algebra
of bounded linear operators on the Hilbert space $\cal H$.
The notion of a quantum state coincides with the notion of a (mathematical) state $\omega$
on $\cal A$. The latter is defined as a linear functional $\omega:\,{\cal A}\mapsto\Co$
which satisfies the conditions of positivity and of normalization
\be
& &\omega(A^*A)\ge 0\quad\mbox{ for all }A\in {\cal A};\\
& &\omega(\Io)=1,
\nnee
where $\Io$ is the identity operator and $A^*$ is the adjoint of $A$.
In particular any state $\omega$ belongs to the dual space of $\cal A$ as a Banach space.

The state $\omega$ is said to be faithful if $\omega(A^*A)=0$ implies $A=0$.

The Gelfand-Naimark-Segal (GNS) construction shows that given a state $\omega$ on a $C^*$-algebra $\cal A$
there exists a *-representation $\pi$
of ${\cal A}$ as bounded linear operators on a Hilbert space ${\cal H}_\omega$, together with an
element $\Omega$ of ${\cal H}_\omega$ such that 
\be
\omega(A)&=&(\pi(A)\Omega,\Omega)\quad \mbox{for all}\quad A\in {\cal A}
\ee
and $\pi_({\cal A})$ is dense in ${\cal H}_\omega$. This representation is unique up to unitary
equivalence. This representation is used here to make the transition from  a situation where quantum states
are described by a  density matrix to the more general context of an arbitrary von Neumann
algebra $\cal A$ of bounded operators on a separable Hilbert space $\cal H$, together with a cyclic
and separating vector $\Omega\in {\cal H}$ of norm one.

In the case of the algebra of all $N$-by-$N$ matrices a simple and explicit realization of
the GNS-representation is possible. See the Appendix.

The relation between a density matrix $\rho$  and the corresponding quantum state $\omega_\rho$, defined by
\be
\omega_\rho(A)&=&\Tr\rho A\quad\mbox{ for all }A\in{\cal A}
\label{repr:omegarho}
\ee
is a one-to-one relation. 
Indeed, if two density
matrices $\rho_1$ and $\rho_2$ produce the same quantum
expectations then they coincide. Conversely, because the Hilbert space $\cal H$ is finite dimensional,
any quantum state $\omega$ determines a density matrix $\rho$ such that $\omega=\omega_\rho$.
The state $\omega_\rho$ is faithful if and only if the density matrix $\rho$ is strictly positive.

For the sake of completeness the proof of the following result is reproduced.

\begin{Theorem}
\label{thm:exists}
Let $\rho$ and $\sigma$ be two strictly positive density matrices operating in a finite dimensional 
Hilbert space $\cal H$. Let $\cal A$ denote the von Neumann algebra of linear operators on $\cal H$.
Let $\pi_\rho,{\cal H}_\rho,\Omega_\rho$ be the GNS-representation induced by $\rho$.
Then there exists a unique strictly positive operator $X$ in the commutant $\pi({\cal A})'$ such that
\be
\Tr\sigma A&=&(\pi_\rho(A)X^{1/2}\Omega_\rho,X^{1/2}\Omega_\rho)
\quad\mbox{ for all }A\in {\cal A}.
\label{finite:exist}
\ee
\end{Theorem}

\beginproof
Because ${\cal H}_\rho$ is finite dimensional and $\Omega_\rho$ is cyclic and separating
one has $\pi_\rho({\cal A})\Omega_\rho=\pi_\rho({\cal A})'\Omega_\rho={\cal H}_\rho$.
Hence, there exists $X$ in $\pi_\rho({\cal A})'$ such that 
\be
X\Omega_\rho&=&\pi_\rho(\sigma\rho^{-1})\Omega_\rho.
\nnee
Then one has for all $A\in {\cal A}$
\be
(\pi_\rho(A)\Omega_\rho,X\Omega_\rho)
&=&
(\pi_\rho(A)\Omega_\rho,\pi_\rho(\sigma\rho^{-1})\Omega_\rho)\cr
&=&
(\pi_\rho(\rho^{-1}\sigma A)\Omega_\rho,\Omega_\rho)\cr
&=&
\Tr\rho \rho^{-1}\sigma A\cr
&=&\Tr\sigma A.
\label{fiite:temp1}
\ee
In particular, take $A=B^*B$ to obtain
\be
(X\pi_\rho(B)\Omega_\rho,\pi_\rho(B)\Omega_\rho)
&=&
\Tr\sigma B^*B\cr
&\ge 0,
\ee
with equality if and only if $B=0$.
This implies that $X$ is a strictly positive operator.

$X\Omega_\rho$ is the unique element of ${\cal H}_\rho$
for which  (\ref {fiite:temp1}) holds.
Because $\Omega_\rho$ is cyclic for $\pi_\rho$ it is separating
for the commutant. Hence, $X$ is unique as well.
\endproof

Introduce the notation ${\cal B}_\rho$ for the real Banach space formed by the
self-adjoint elements $K$ of $\pi_\rho({\cal A})'$
satisfying $(K\Omega_\rho,\Omega_\rho)=0$.

\begin{Theorem}
\label {theorem2}
Let $\cal H$, $\cal A$ and $\rho$ be as in the previous Theorem.
Let $\pi_\rho,{\cal H}_\rho,\Omega_\rho$  the GNS-representation induced by $\rho$.
There is a one-to-one correspondence $\xi_\rho$ between faithful states $\omega$ on $\cal A$
and elements of ${\cal B}_\rho$. It satisfies
\be
\omega(A)&=&e^{-\alpha_\rho(K)}(\pi_\rho(A)e^{\frac 12 K}\Omega_\rho,e^{\frac 12 K}\Omega_\rho)
\quad\mbox{ for all }A\in {\cal A},
\label{finite:chart}
\ee
with $K=\xi_\rho(\omega)$ and the function $\alpha_\rho$ given by
\be
\alpha_\rho(K)&=&\log (e^{K}\Omega_\rho,\Omega_\rho).
\nnee
\end{Theorem}

\beginproof
Let $\omega=\omega_\sigma$.
The previous theorem guarantees the existence of a unique strictly positive operator $X$
in the commutant $\pi_\rho({\cal A})'$. This operator $X$ can be exponentiated.
Let
\be
K=\log X-(\log X\Omega_\rho,\Omega_\rho).
\nnee
Then $(K\Omega_\rho,\Omega_\rho)=0$ holds by construction
and (\ref {finite:chart}) is satisfied with $\alpha_\rho(K)=-(\log X\Omega_\rho,\Omega_\rho)$
(remember that $(X\Omega_\rho,\Omega_\rho)=1$).

Conversely, given $K$, the r.h.s.~of (\ref {finite:chart}) defines a faithful state $\omega$ of $\cal A$.
\endproof

The map $\xi_\rho$ is a chart which makes the manifold $\Mo$
of all faithful quantum states into a Banach manifold.
The chart $\xi_\rho$ is said to be {\em centered} at $\rho$. 
It satisfies $\xi_\rho(\rho)=0$.

All representations $\pi_\rho$, ${\cal H}_\rho$, with $\rho$ strictly positive,
are unitary equivalent and can be identified.
Therefore, in what follows the index $\rho$ of $\pi_\rho$ is dropped 
and the Hilbert space in which the representation $\pi$
works is denoted ${\cal H}_\pi$. 


\section{The tangent plane at the center}
\label{sect:tangent}

Let $K=\xi_\rho(\sigma)\in{\cal B}_\rho$.
Introduce the notation
\be
\Psi_K&=&e^{\frac  12(K-\alpha(K))}\Omega_\rho.
\nnee
One has
\be
\frac{\upd\,}{\upd t}\bigg|_{t=0}
\alpha_\rho(tK)
&=&\frac{(Ke^{tK}\Omega_\rho,\Omega_\rho)}{(e^{tK}\Omega_\rho,\Omega_\rho)}\bigg|_{t=0}\cr
&=&
(K\Omega_\rho,\Omega_\rho)\cr
&=&0
\ee
and
\be
\frac{\upd\,}{\upd t}\bigg|_{t=0}
\Psi_{tK}
&=&
\frac{\upd\,}{\upd t}\bigg|_{t=0}
e^{\frac  12(tK-\alpha(tK))}\Omega_\rho\cr
&=&
\frac 12 K\Omega_\rho.
\nnee
The density matrix $\sigma_t$, defined by
\be
\Tr\sigma_t A&=&(\pi(A)\Psi_{tK},\Psi_{tK}),
\quad A\in {\cal A},
\nnee
satisfies
\be
\frac{\upd\,}{\upd t}\bigg|_{t=0}
\Tr\sigma_t A
&=&
(\pi(A)K\Omega_\rho,\Omega_\rho).
\nnee
Hence the linear functional $f_{\rho,K}$ defined by
\be
A\in {\cal A}\mapsto f_{\rho,K}(A)=(\pi(A)\Omega_\rho,K\Omega_\rho).
\label {tangent:fk}
\ee
is the derivative of the quantum state $\omega_\rho$ 
in the direction $\omega_\sigma$, where the density matrix $\sigma=\sigma_1$ has the property
that $\xi_\rho(\sigma)=K$.

One concludes that the tangent plane $\To_\rho\Mo$ at the point $\omega_\rho\in\Mo$
consists of all linear hermitian functionals $f_{\rho,K}:\,{\cal A}\mapsto\Co$
of the form (\ref {tangent:fk}), with $K\in{\cal B}_\rho$.
The functional $f_{\rho,K}$ belongs to the dual of $\cal A$.
In addition,
\be
||f_{\rho,K}||
&=&\sup_{A\in{\cal A}}\left\{f_{\rho,K}(A):\,||A||\le 1\right\}\cr
&=&\sup_{A\in{\cal A}}\left\{(\pi(A)K\Omega_\rho,\Omega_\rho):\,||A||\le 1\right\}\cr
&=&||K^{1/2}\Omega_\rho||^2\cr
&\le&||K^{1/2}||^2=||K||.
\nnee
Hence, $K\mapsto f_{\rho,K}$ is a bounded linear operator.
This is a prerequisite for proving
in the next Theorem that this map is the Fr\'echet derivative of the chart $\xi_\rho$.
This bounded operator is denoted $F_\rho$ in what follows. One has $F_\rho K=f_{\rho,K}$.
The inverse operator $F_\rho^{-1}$ satisfies $F_\rho^{-1}f_{\rho,K}=K$. It is well-defined.
Indeed, $f_{\rho,K}=f_{\rho,L}$ implies for all $B\in {\cal A}$
\be
0&=&(\pi(B^*B)(K-L)\Omega_\rho,\Omega_\rho)\cr
&=&(K-L)\pi(B)\Omega_\rho,\pi(B)\Omega_\rho).
\nnee
Because $\Omega_\rho$ is a cyclic vector there follows that $K=L$.

\begin{Theorem}
\label{theorem:frechet}
The inverse of the map $\xi_\rho:\,\Mo\mapsto {\cal B}_\rho$, defined in Theorem \ref {theorem2},
is Fr\'echet-differentiable at $\omega=\omega_\rho$.
The Fr\'echet derivative is denoted $F_\rho$. It maps $K$ to $f_{\rho,K}$, where the 
latter is defined by (\ref {tangent:fk}).
\end{Theorem}

\beginproof

Let $K=\xi_\rho(\omega_\sigma)$. One calculates
\be
||\omega_\sigma-\omega_\rho-F_\rho K||
&=&
\sup_{A\in{\cal A}}\left\{|\omega_\sigma(A)-\omega_\rho(A)-F_\rho K(A)|:\, ||A||\le 1\right\}\cr
&=&
\sup_{A\in{\cal A}}\left\{
|(\pi(A)\Omega_\rho,[e^{K-\alpha(K)}-\Io-K]\Omega_\rho)|:\, ||A||\le 1\right\}\cr
&\le&
||e^{K-\alpha(K)}-\Io-K||\cr
&\le&
|\alpha(K)|+\mbox{ o}(||K-\alpha(K)||).
\label{tangent:temp2}
\ee
Note that
\be
|\alpha(K)|\le\log||e^K||\le||K||
\nnee
and
\be
||K-\alpha(K)||\le 2||K||.
\ee
One concludes that (\ref {tangent:temp2}) converges to 0 as $||K||$ tends to 0.
This proves that $F_\rho K$ is the Fr\'echet derivative of
$\xi_\rho(\omega_\sigma)\mapsto \omega_\sigma$ at $\sigma=\rho$.
\endproof


\section{The atlas}
\label{sect:atlas}

Following the approach of Pistone and collaborators \cite{PS95,PR99,PG13,MP17}
we build an atlas of charts $\xi_\rho$, one for each strictly positive density matrix $\rho$.
The compatibility of the different charts requires the study of the cross-over map
$\xi_{\rho_1}(\omega_\sigma)\mapsto \xi_{\rho_2}(\omega_\sigma)$,
where $\rho_1,\rho_2,\sigma$ are arbitrary strictly positive density matrices.

Simplify notations by writing $\xi_1$ and $\xi_2$
instead of $\xi_{\rho_1}$, respectively $\xi_{\rho_2}$. Similarly, write $\Omega_1$ and $\Omega_2$
instead of $\Omega_{\rho_1}$, respectively $\Omega_{\rho_2}$,
and $F_1,F_2$ instead of $F_{\rho_1}$, respectively $F_{\rho_2}$.


\begin{Proposition}
\label{atlas:crossover}
The cross-over map $\xi_2\circ\xi_1^{-1}$ equals the linear operator $F^{-1}_2F_1$.
\end{Proposition}

\beginproof
Let $\omega_\sigma=\xi_1^{-1}K$ and $L=\xi_2\circ\xi_1^{-1}K$. Then 
\be
\xi_1(\omega_\sigma)=K
\quad\mbox{ and }
\xi_2(\omega_\sigma)=L.
\label{atlas:temp3}
\nnee
of the inverse of the map $\omega_\sigma\mapsto\xi_1(\sigma)$ and $F_2$ is the Fr\'echet derivative
of the inverse of the map $\omega_\sigma\mapsto\xi_2(\sigma)$.
Hence one has $F_1K=f_{1,K}$
and $F_2L=f_{2,L}$. From (\ref {atlas:temp3}) then follows that
\be
f_{2,L}(A)
&=&(\pi(A)\Omega_2,L\Omega_2)\cr
&=&\omega_\sigma(A)\cr
&=&(\pi(A)\Omega_1,K\Omega_1)\cr
&=&f_{1,K}(A).
\nnee
One concludes that $F_2L=f_{2,L}=f_{1,K}=F_1K$ and hence $L=F_2^{-1}F_1K$.
In combination with $L=\xi_2\circ\xi_1^{-1}K$ this proves the proposition.
\endproof

Continuity of the cross-over map follows from the continuity of the exponential and logarithmic functions and from
the following estimate. By the previous result this implies that also the linear operator $F^{-1}_2F_1$
is continuous. The latter is of course a trivial statement because of the assumption of finite dimensions.

\begin{Proposition}
For any pair $\sigma_1,\sigma_2$ of strictly positive density matrices is
\be
||e^{\xi_2(\sigma_2)}-e^{\xi_2(\sigma_1)}||&\le&
||\rho_1||\,||\rho_2^{-1}||\,||e^{\xi_1(\sigma_2)}-e^{\xi_1(\sigma_1)}||.
\nnee
\end{Proposition}

\beginproof
Let $\xi_i(\omega_j)=K_{i,j}$. 
One calculates
\be
||e^{K_{2,2}}-e^{K_{2,1}}||^2
&=&
\sup_{A\in{\cal A}, A\not=0}\frac{\left|\left(\left[e^{K_{2,2}}-e^{K_{2,1}}\right]\pi(A)\Omega_2,\pi(A)\Omega_2\right)\right|}
{||\pi(A)\Omega_2||^2}\cr
&=&
\sup_{A\in{\cal A}, A\not=0}\frac{\left|\Tr(\sigma_2-\sigma_1)A^*A\right|}{\Tr\rho_2 A^*A}.
\nnee
Note that 
\be
||\rho_1||\,||\rho_2^{-1}||\,\Tr\rho_2 A^*A
\ge ||\rho_1||\,\Tr A^*A
\ge \Tr\rho_1 A^*A
\nnee
Use this to obtain
\be
||e^{K_{2,2}}-e^{K_{2,1}}||^2
&\le&
||\rho_1||\,||\rho_2^{-1}||\,
\sup_{A\in{\cal A}, A\not=0}\frac{\left|\Tr(\sigma_2-\sigma_1)A^*A\right|}{\Tr\rho_1 A^*A}\cr
&=&
||\rho_1||\,||\rho_2^{-1}||\,||e^{K_{1,2}}-e^{K_{1,1}}||^2.
\nnee
\endproof

The Fr\'echet derivative of the bounded linear operator $F^{-1}_2F_1$ is the operator itself.
From Proposition \ref {atlas:crossover} follows therefore that the cross-over map $\xi_2\circ\xi_1^{-1}$
is Fr\'echet differentiable. Hence, the following theorem holds.

\begin{Theorem}
\label{atlas:theorem}
The set $\Mo$ of faithful states on the algebra $\cal A$ of square matrices,
together with the atlas of charts $\xi_\rho$,
where $\xi_\rho$ is defined by Theorem \ref{thm:exists},
is a differentiable Banach manifold. For any pair of strictly positive density matrices
$\rho$ and $\sigma$ the cross-over map $\xi_\sigma\circ\xi_\rho^{-1}$ is a bounded linear operator.
\end{Theorem}

\section{The Bogoliubov inner product}
\label{sect:bogo}

Umegaki's divergence/relative entropy $D(\sigma,\tau)$ of a pair of strictly positive density
matrices $\sigma$ and $\tau$ is defined by \cite{UH62,LG75,AH76}
\be
D(\sigma||\tau)&=&\Tr\sigma(\log\sigma-\log\tau).
\nnee
It can be used to define a metric tensor $g_{\sigma,\tau}(\rho)$, as explained below.

Introduce $\sigma_s$ and $\tau_t$ given by
\be
\sigma_s&=&\frac{1}{Z(s)}\exp\left(\log\rho+s(\log\sigma-\log\rho)\right),
\label{bog:sigmas}\\
\tau_t&=&\frac{1}{W(t)}\exp\left(\log\rho+t(\log\tau-\log\rho)\right)
\nnee
with
\be
Z(s)&=&\Tr \exp\left(\log\rho+s(\log\sigma-\log\rho)\right),\\
W(t)&=&\Tr \exp\left(\log\rho+t(\log\tau-\log\rho)\right).
\nnee
Both $\sigma_s$ and $\tau_t$ are well-defined density matrices.
The maps $s\mapsto\omega_{\sigma_s}$ and $t\mapsto\omega_{\tau_t}$ describe two orbits in $\Mo$,
intersecting at $\omega_\rho$:
$\sigma_0=\tau_0=\rho$. For further use, note that $Z(0)=W(0)=Z(1)=W(1)=1$.

The metric tensor $g_{\sigma,\tau}(\rho)$ is defined by
\be
g_{\sigma,\tau}(\rho)&=&-\frac{\partial\,}{\partial s}\frac{\partial\,}{\partial t}\bigg|_{s=t=0}
D(\sigma_s||\tau_t).
\label{bog:defg}
\ee
With the help of the identity
\be
\frac{\upd\,}{\upd t}\bigg|_{t=0}e^{H+tA}
&=&
\int_0^1\upd u\,e^{uH}Ae^{(1-u)H}
\nnee
one obtains
\be
\frac{\upd\,}{\upd t}\bigg|_{t=0}\log \tau_t
&=&
\log\tau-\log\rho-\frac{\upd\,}{\upd t}\bigg|_{t=0}\log W(t)\cr
&=&
\log\tau-\log\rho-\int_0^1\upd u\, \Tr \rho^u(\log\tau-\log\rho)\rho^{1-u}\cr
&=&
\log\tau-\log\rho+D(\rho||\tau)
\nnee
so that
\be
g_{\sigma,\tau}(\rho)
&=&
\frac{\partial\,}{\partial s}\frac{\partial\,}{\partial t}\bigg|_{s=t=0}
\Tr\sigma_s\log\tau_t\cr
&=&
\frac{\upd \,}{\upd s}\bigg|_{s=0}
\Tr\sigma_s\left[\log \tau-\log\rho+D(\rho||\tau)\right]\cr
&=&
\frac{\upd \,}{\upd s}\bigg|_{s=0}
\Tr\sigma_s\left[\log \tau-\log\rho\right]\cr
&=&
\int_0^1\upd u\, \Tr \rho^u(\log\sigma-\log\rho)\rho^{1-u}
\left[\log \tau-\log\rho)\right]\cr
& &
-\left(\frac{\upd \,}{\upd s}\bigg|_{s=0}Z(s)\right)
\Tr\rho\left[\log \tau-\log\rho\right]\cr
&=&
\int_0^1\upd u\, \Tr \rho^u(\log\sigma-\log\rho)
\rho^{1-u}(\log\tau-\log\rho)\cr
& &-D(\rho||\sigma)D(\rho||\tau).
\label{kubomori:inner}
\ee
This is the inner product of Bogoliubov.
Its positivity is shown in the next section.
It is straightforward to check that $g_{\sigma,\tau}=g_{\tau,\sigma}$.


\section{Alternative charts}
\label{sect:alt}

The inner product (\ref {kubomori:inner}) is expressed
in terms of density matrices rather than tangent vectors.
Let us therefore calculate the tangent vector of the orbit $\omega_{\sigma_s}$
defined by (\ref {bog:sigmas}).

\begin{Lemma}
\label{alt:lemma1}
For each $A$, self-adjoint element of $\cal A$ such that $\omega_\rho(A)=0$,
there exists a unique element $K$ of ${\cal B}_\rho$ such that
\be
\int_0^1\upd u\,\pi\left(\rho^{u} A\rho^{-u}\right)\Omega_\rho
&=&K\Omega_\rho.
\label{alt:exists}
\ee
\end{Lemma}

\beginproof
An operator $K$ in the commutant $\pi({\cal A})'$ satisfying (\ref{alt:exists}) exists
because $\pi({\cal A})'\Omega_\rho={\cal H}_\pi$. It is unique because $\Omega_\rho$ is
separating for $\pi({\cal A})'$.
It satisfies
\be
(K\Omega_\rho,\Omega_\rho)
&=&
\int_0^1\upd u\,\pi\left(\rho^{u} A\rho^{-u}\right)\Omega_\rho,\Omega_\rho)\cr
&=&
\int_0^1\upd u\,\Tr\rho^{1+u} A\rho^{-u}\cr
&=&\omega_\rho(A)\cr
&=&0.
\nnee

Finally, for any $B$ in $\cal A$ is
\be
(K^*\Omega_\rho,\pi(B)\Omega_\rho)
&=&
(\pi(B^*)\Omega_\rho,K\Omega_\rho)\cr
&=&
(\pi(B^*)\Omega_\rho,\int_0^1\upd u\,\pi\left(\rho^{u} A\rho^{-u}\right)\Omega_\rho)\cr
&=&
\int_0^1\upd u\,(\pi(\rho^{-u} A\rho^{u}B^*)\Omega_\rho,\Omega_\rho)\cr
&=&
\int_0^1\upd u\,\Tr\rho^{1-u}A\rho^{u}B^*\cr
&=&
\int_0^1\upd u\,\Tr\rho^{u}A\rho^{1-u}B^*\cr
&=&
\int_0^1\upd u\,(\pi(B^*\rho^u A\rho^{-u})\Omega_\rho,\Omega_\rho)\cr
&=&(K\Omega_\rho,\pi(B)\Omega_\rho).
\nnee
This shows that $K=K*$. One concludes that $K$ belongs to ${\cal B}_\rho$.
\endproof

\begin{Lemma}
\label{alt:lemma3}
There exists a strictly positive operator $G_\rho$ on ${\cal H}_\pi$
which satisfies
\be
G_\rho \int_0^1\upd u\,\pi\left(\rho^uA\rho^{-u}\right)\Omega_\rho&=&\pi(A)\Omega_\rho
\quad\mbox{ for all }\quad A\in {\cal A}.
\nnee
\end{Lemma}

\beginproof
First consider the operator $X$ defined by
\be
X\pi(A)\Omega_\rho
&=&
\int_0^1\upd u\,\pi\left(\rho^uA\rho^{-u}\right)\Omega_\rho.
\nnee
It is well-defined because $\pi(A)\Omega_\rho=0$ implies $A=0$.
It is a positive operator. This follows from
\be
(X\pi(A)\Omega_\rho,\pi(A)\Omega_\rho)
&=&
\int_0^1\upd u\,\pi\left(\left(\rho^uA\rho^{-u}\right)\Omega_\rho,\pi(A)\Omega_\rho)\right)\cr
&=&
\int_0^1\upd u\,\Tr\rho A^*\rho^uA\rho^{-u}\cr
&=&
\int_0^1\upd u\,\Tr\rho^{(1-u)/2}A^*\rho^uA\rho^{(1-u)/2}\cr
&\ge& 0.
\nnee
The latter expression vanishes if and only if $\rho^{(1-u)/2}A^*\rho^uA\rho^{(1-u)/2}=0$
for almost all $u$ in $[0,1]$. Because $\rho$ is strictly positive this can happen only if $A=0$.
This shows that the operator $X$ is invertible.
Take $G_\rho$ equal to the inverse of $X$ to obtain the desired result.
\endproof

\begin{Theorem}
\label {theorem3}
Let $\cal H$, $\cal A$ and $\rho$ be as in the previous theorems.
Let $\pi,{\cal H}_\pi,\Omega_\rho$ be the GNS-representation of $\cal A$ induced by $\rho$.
Let $G_\rho$ be the positive operator defined by the previous lemma.
\begin{description}
 \item i) There exists a map $\chi_\rho$ from the faithful states $\omega$ on $\cal A$
into the real Banach space ${\cal B}_\rho$, formed by the
self-adjoint elements $K$ of $\pi_\rho({\cal A})'$
satisfying $(K\Omega_\rho,\Omega_\rho)=0$,
such that for any strictly positive density matrix $\sigma$ one has
\be
G_\rho\chi_\rho(\omega_\sigma)\Omega_\rho&=&
\pi(A_{\rho,\sigma})\Omega_\rho.
\nnee
with $A_{\rho,\sigma}$ in $\cal A$ given by
\be
A_{\rho,\sigma} &=&
\log\sigma-\log\rho+D(\rho||\sigma).
\nnee
\item ii) The map $\chi_\rho$ is injective.
\item iii) For each strictly positive density matrix $\sigma$
is
\be
\frac{\upd \,}{\upd s}\bigg|_{s=0}\omega_{\sigma_s}
&=&f_{\rho,K},
\nnee
with $K=\chi_\rho(\omega_\sigma)$,
where $\sigma_s$ is defined by (\ref {bog:sigmas})
and $f_{\rho,K}$ is defined by (\ref {tangent:fk}).
\end{description}

\end{Theorem}

\beginproof

\paragraph{i)}
By the previous lemma one has
$\pi(A_{\rho,\sigma})\Omega_\rho=G_\rho X\Omega_\rho$ with $X$ defined by
\be
X
&=&
\int_0^1\upd u\,\pi\left(\rho^u A_{\rho,\sigma}\rho^{-u}\right).
\nnee
Note that $A_{\rho,\sigma}$ is self-adjoint and satisfies $\omega_\rho (A_{\rho,\sigma})=0$.
Hence, by Lemma \ref {alt:lemma1} there exists a unique $K$ in ${\cal B}_\rho$ such that
$X\Omega_\rho=K\Omega_\rho$.
This shows that the map $\chi_\rho$ which maps $\omega_\sigma$ onto this element $K$ of ${\cal B}_\rho$
is well-defined.

\paragraph{ii)}
Assume that $\chi_\rho(\omega_\sigma)=\chi_\rho(\omega_\tau)$. This implies
$A_{\rho,\sigma}=A_{\rho,\tau}$ and hence
\be
\log\sigma+D(\rho||\sigma)&=&\log\tau+D(\rho||\tau).
\nnee
The latter implies
\be
\sigma &=&\tau e^{D(\rho||\tau)-D(\rho||\sigma)}.
\nnee
Because $\Tr\sigma=\Tr\tau$ there follows that $D(\rho||\tau)=D(\rho||\sigma)$
and hence $\sigma=\tau$. This shows that the map $\chi_\rho$ is injective.

\paragraph{iii)}
One has for all $B\in{\cal A}$
\be
\frac{\upd \,}{\upd s}\bigg|_{s=0}
\Tr\sigma_s B
&=&
D(\rho||\sigma)\omega_\rho(B)
+\int_0^1\upd u\,\Tr \rho^{1-u}(\log\sigma-\log\rho)\rho^{u} B\cr
&=&\int_0^1\upd u\,\Tr \rho^{1-u}A_{\rho,\sigma}\rho^{u} B\cr
&=&\int_0^1\upd u\,\Tr \rho^{1-u} B \rho^{u}A_{\rho,\sigma}\cr
&=&\int_0^1\upd u\,\left(\pi\left(B\rho^{u}A_{\rho,\sigma}\rho^{u}\right)\Omega_\rho,\Omega_\rho\right)\cr
&=&
\left(\pi(B)G^{-1}_{\rho}\pi(A_{\rho,\sigma})\Omega_\rho,\Omega_\rho\right)\cr
&=&
\left(\pi(B)\chi_\rho(\omega_\sigma)\Omega_\rho,\Omega_\rho\right)\cr
&=&f_{\rho,K}(B),
\nnee
with $K=\chi_\rho(\omega_\sigma)$.
\endproof

\section{The Riemannian metric}
\label{sect:riem}

Introduce an inner product $\langle\cdot,\cdot\rangle_\rho$ on $T_\rho\Mo$ defined by
\be
\langle f_{\rho,P}, f_{\rho,Q}\rangle_\rho
&=&
\langle G_\rho P\Omega_\rho,Q\Omega_\rho),
\label{riem:innprod}
\ee
for any $P,Q$ in ${\cal B}_\rho$.
The matrix $G_\rho$, which is introduced in Lemma \ref {alt:lemma3}, is strictly positive.
Hence the inner product is positive and non-degenerate.
It defines a Riemannian geometry on the manifold $\Mo$.

\begin{Theorem}
Let $\rho,\sigma,\tau$ be strictly positive density matrices.
Let $\chi_\rho$ be the chart defined in Theorem \ref {theorem3}.
Let $\langle\cdot,\cdot\rangle_\rho$ be the inner product defined on the tangent plane $T_\rho\Mo$
by (\ref {riem:innprod}).
The inner product of Bogoliubov, defined by (\ref {bog:defg}), satisfies
\be
g_{\sigma,\tau}(\rho)&=&\langle f_{\rho,P}, f_{\rho,Q}\rangle_\rho,
\nnee 
with $P=\chi_\rho(\sigma)$ and $Q=\chi_\rho(\tau)$.
\end{Theorem}

\beginproof
From (\ref {kubomori:inner}) follows
\be
g_{\sigma,\tau}(\rho)
&=&
\int_0^1\upd u\,\left(\pi\left(\rho^{-u}(\log\tau-\log\rho)\rho^u(\log\sigma-\log\rho)\right)\Omega_\rho,
\Omega_\rho\right)\cr
& &
-D(\rho||\sigma)D(\rho||\tau)\cr
&=&
\int_0^1\upd u\,
\left(\pi\left(\rho^{-u}(A_{\rho,\tau}-D(\rho||\tau))\rho^u(A_{\rho,\sigma}-D(\rho||\sigma))\right)\Omega_\rho,
\Omega_\rho\right)\cr
& &
-D(\rho||\sigma)D(\rho||\tau).
\nnee
Use now that
\be
(\pi(A_{\rho,\sigma})\Omega_\rho,\Omega_\rho)
=\int_0^1\upd u\,\left(\pi(\rho^{-u}A_{\rho,\tau}\rho^{u})\Omega_\rho,\Omega_\rho\right)=0
\nnee
to obtain
\be
g_{\sigma,\tau}(\rho)
&=&
\int_0^1\upd u\,
\left(\pi\left(\rho^{-u}(A_{\rho,\tau})\rho^u(A_{\rho,\sigma})\right)\Omega_\rho,
\Omega_\rho\right).
\nnee
This can be written as
\be
g_{\sigma,\tau}(\rho)
&=&
\left(\pi\left(A_{\rho,\sigma})\right)\Omega_\rho,G^{-1}_\rho \pi(A_{\rho,\tau})\Omega_\rho\right)\cr
&=&
\left(G_\rho\chi_\rho(\sigma)\Omega_\rho,\chi_\rho(\tau)\Omega_\rho\right)\cr
&=&
\langle f_{\rho,P}, f_{\rho,Q}\rangle_\rho.
\nnee
\endproof


\section{The mixture connection}
\label{sect:mix}

Consider the situation in which the affine combinations of the form
\be
t\mapsto\rho_t=(1-t)\rho_0+t\rho_1
\nnee
are the geodesics of the geometry.
Introduce the abbreviation $\omega_t=\omega_{\rho_t}$, with the latter defined by (\ref {repr:omegarho}).
The derivative
\be
\frac{\upd\,}{\upd t}\omega_t&=&\omega_1-\omega_0
\nnee
is a tangent vector which is constant. This implies a vanishing connection.


\section{The exponential connection}
\label{sect:exp}

On the other hand, in the case of the exponential connection the geodesics 
$t\mapsto\rho_t$ are such that
\be
\log\rho_t&=&(1-t)\log\rho_0+t\log\rho_1-\zeta(t)\cr
&=&
\log\rho_0+tH-\zeta(t),
\label{exp:def}
\ee
where $\zeta(t)$ is a normalizing function and $H$ is defined by $H=\log\rho_1-\log\rho_0$.
Note that
\be
\omega_t(H)&=&\Tr\rho_t H\cr
&=&\Tr\rho_t\left[\log\rho_1-\log\rho_0\right]\cr
&=&D(\rho_t||\rho_0)-D(\rho_t||\rho_1).
\label{exp:Hav}
\ee

One has
\be
\frac{\upd\,}{\upd t}\rho_t
&=&\frac{\upd\,}{\upd \epsilon}\bigg|_{\epsilon=0}
\exp(\log\rho_t+\epsilon H)
-\frac{\upd\zeta}{\upd t}\rho_t\cr
&=&
\int_0^1\upd u\,(\rho_t)^uH(\rho_t)^{1-u}
-\frac{\upd\zeta}{\upd t}\rho_t.
\nnee
Therefore, the derivative of the quantum state becomes
\be
\frac{\upd\,}{\upd t}\omega_t(A)
&=&
\int_0^1\upd u\,\Tr(\rho_t)^uH(\rho_t)^{1-u}A\cr
& &-\frac{\upd\zeta}{\upd t}\omega_t(A).
\label{exp:der}
\ee
Take $A=\Io$ to find that
\be
\frac{\upd\zeta}{\upd t}
=
\omega_t(H).
\label{exp:alphadot}
\ee

\begin{Proposition}
The function $\zeta(t)$ is convex.
\end{Proposition}

\beginproof
Let $A=\rho_t^{u/2} H\rho_t^{-u/2}$. Then one has
\be
\Tr\rho_t^{1-u}H\rho_t^u H&=&\Tr\rho_t A^*A\cr
&\ge&|\Tr\rho_t A|^2\cr
&=&|\omega_t(H)|^2.
\nnee
Use this in
\be
\frac{\upd^2\,}{\upd t^2}\zeta(t)
&=&
\frac{\upd\,}{\upd t}\Tr \rho_t H\cr
&=&
\int_0^1\upd u\,\left[\Tr\rho_t^{1-u}H\rho_t^u H-|\omega_t(H)|^2\right]\cr
&\ge&0.
\label{exp:secder}
\ee
\endproof

Because $\zeta(0)=\zeta(1)=0$ there follows that $\zeta(t)\le 0$ on $0<t<1$.
From (\ref {exp:Hav}) and 
\be
\frac{\upd\,}{\upd t}\omega_t(H)=\frac{\upd^2\,}{\upd t^2}\zeta(t)\ge 0
\nnee
follows that the expectation $\omega_t(H)$ increases from $\omega_0(H)=-D(\rho_0||\rho_1)\le 0$
to $\omega_1(H)=D(\rho_1||\rho_0)\ge 0$.

\begin{Proposition}
\label{exp:prop2}
$\upd\omega_t/\upd t$ is the derivative of $\omega_t$
in the direction $\omega$, with $\omega$ such that
$\xi_t(\omega)=\chi_t(\rho_1)-\chi_t(\rho_0)$.
\end{Proposition}

\beginproof
From the definitions of $\chi_\rho$, $G_\rho$ and $A_{\sigma,\tau}$ follows
\be
(\pi(A)\Omega_t,\chi_t(\omega_\sigma)\Omega_t)
&=&
(\pi(A)\Omega_t,G^{-1}_t\pi(A_{\sigma_t,\sigma})\Omega_t)\cr
&=&
\int_0^1\upd u\,\left(\pi(A)\Omega_t,\pi(\rho^u_t A_{\sigma_t,\sigma}\rho^{-u}_t)\Omega_t\right)\cr
&=&
\int_0^1\upd u\,\left(\pi(\rho^{-u}_t A_{\sigma_t,\sigma}\rho^{u}_tA)\Omega_t,\Omega_t\right)\cr
&=&D(\rho_t||\sigma)\rho_t(A)\cr
& &
+\int_0^1\upd u\,\left(\pi(A)\Omega_t,
\pi\left(\rho_t^u(\log\sigma-\log\rho_t)\rho_t^{-u}\right)
\Omega_t\right).
\nnee
Use this for $\sigma=\rho_1$ and for $\sigma=\rho_0$ and subtract.
This gives
\be
(\pi(A)\Omega_t,[\chi_t(\rho_1)-\chi_t(\rho_0]\Omega_t)
&=&\left[D(\rho_t||\rho_1)-D(\rho_t||\rho_0)\right]\rho_t(A)\cr
&+&
\int_0^1\upd u\,\left(\pi(A)\Omega_t,
\pi\left(\rho_t^u(\log\rho_1-\log\rho_0)\rho_t^{-u}\right)
\Omega_t\right)\cr
&=&-\omega_t(H)\rho_t(A)\cr
&+&
\int_0^1\upd u\,\left(\pi(A)\Omega_t,
\pi\left(\rho_t^uH\rho_t^{-u}\right)
\Omega_t\right)\cr
&=&-\omega_t(H)\rho_t(A)\cr
&+&
\int_0^1\upd u\,\Tr \rho_t^{1-u}H\rho_t^u A\cr
&=&
\frac{\upd\,}{\upd t}\omega_t(A)\cr
&=&f_{t,K},
\nnee
with $K=\chi_t(\rho_1)-\chi_t(\rho_0)$.
\endproof

The following result shows that $\chi_\rho(\omega)$ is an affine coordinate
in the case of the exponential connection.

\begin{Proposition}
\label{exp:prop3}
Let be given a strictly positive density matrix $\rho$ and a geodesic $t\mapsto \rho_t$
of the form (\ref {exp:def}). Then
\be
\chi_\rho(\omega_t)&=&(1-t)\chi_\rho(\omega_0)+t\chi_\rho(\omega_1).
\label{exp:aff}
\ee
\end{Proposition}

\beginproof
From iii) of Theorem \ref {theorem3} follows that for all $A\in{\cal A}$
\be
(\pi(A)\Omega_\rho,\chi_\rho(\omega_t))
&=&\frac{\upd\,}{\upd s}\bigg|_{s=0}\Tr\sigma_s A,
\nnee
with $\sigma=\rho_t$.
The latter implies
\be
\sigma_s&=&\frac{1}{Z_t(s)}\exp\left(\log\rho+s(\log\rho_t-\log\rho)\right),
\nnee
where
\be
Z_t(s)&=&\Tr \exp\left(\log\rho+s(\log\rho_t-\log\rho)\right).
\nnee
Use (\ref {exp:def}) to obtain 
\be
\sigma_s&=&\frac{1}{Z_t(s)}\exp\left((1-s)\log\rho+s((1-t)\log\rho_0+t\log\rho_1-\zeta(t))\right).
\nnee
There follows
\be
(\pi(A)\Omega_\rho,\chi_\rho(\omega_t))
&=&
\int_0^1\upd u\,\Tr \rho^u\left[-\log\rho+(1-t)\log\rho_0+t\log\rho_1-\zeta(t)\right]\rho^{1-u}A\cr
& &-\frac{\upd\,}{\upd s}\bigg|_{s=0}Z_t(s)\omega_\rho(A)\cr
&=&
(1-t)\int_0^1\upd u\,\Tr \rho^u\left[-\log\rho+\log\rho_0-\zeta(t)\right]\rho^{1-u}A\cr
& &
+t\int_0^1\upd u\,\Tr \rho^u\left[-\log\rho+\log\rho_1-\zeta(t)\right]\rho^{1-u}A\cr
& &
-\frac{\upd\,}{\upd s}\bigg|_{s=0}Z_t(s)\omega_\rho(A)\cr
&=&
(1-t)(\pi(A)\Omega_\rho,\chi_\rho(\omega_0))+t(\pi(A)\Omega_\rho,\chi_\rho(\omega_1))\cr
& &
\frac{\upd\,}{\upd s}\bigg|_{s=0}Z_0(s)\omega_\rho(A)+
\frac{\upd\,}{\upd s}\bigg|_{s=0}Z_1(s)\omega_\rho(A)\cr
& &
-\frac{\upd\,}{\upd s}\bigg|_{s=0}Z_t(s)\omega_\rho(A)\cr
&=&
\left(\pi(A)\Omega_\rho,(1-t)\chi_\rho(\omega_0)+t\chi_\rho(\omega_1)\right).
\nnee
Because $A\in{\cal A}$ is arbitrary and $\Omega_\rho$ is cyclic and separating one concludes
(\ref {exp:aff}).
\endproof


\section{Modular automorphisms}
\label{sect:modul}

The quantum manifold $\Mo$ carries an additional structure, which is induced by the modular automorphism groups,
one for each $\rho\in \Mo$. In the commutative case the automorphisms become trivial.

The Tomita-Takesaki theory \cite{TM70} associates
with each state $\omega_\rho$ in $\Mo$ a self-adjoint operator $\Delta_\rho$ on ${\cal H}_\pi$,
called the modular operator. The one-parameter group of unitary operators $\Delta_\rho^{it}$
defines a group of inner automorphisms of the algebra $\cal A$. Indeed, for any $A$ in $\cal A$
the operator $\Delta_\rho^{it}\pi(A)\Delta_\rho^{-it}$ belongs again to $\pi({\cal A})$.
In particular, it induces a group of transformations of the manifold $\Mo$ by mapping any
state $\omega_\sigma$ onto the state $\omega_{\sigma,t}$ defined by
\be
\omega_{\sigma,t}(A)&=&\left(\Delta_\rho^{it}\pi(A)\Delta_\rho^{-it}\Omega_\sigma,\Omega_\sigma\right).
\nnee
This group of transformations has $\omega_\rho$ as a fixpoint because $\Delta\Omega_\rho=\Omega_\rho$.

A useful property of the group of modular automorphims is the so-called {\em KMS condition},
named after Kubo, Martin and Schwinger.
Given two elements $A$ and $B$ of $\cal A$ the function $F(t)$ defined by
\be
F(t)&=&\left(\pi(A)\Delta_\rho^{it}\pi(B)\Omega_\rho,\Omega_\rho\right)
\nnee
has an analytic continuation in the complex plane such that
\be
F(t+i)&=&\left(\pi(B)\Delta_\rho^{it}\pi(A)\Omega_\rho,\Omega_\rho\right).
\nnee
This property captures the essence of cyclic permutation under the trace
and is helpful in the more general context when manipulating non-commuting pairs of operators.


\section{Discussion}

This paper reviews known and less known results of quantum information geometry.
The Hilbert space is assumed to be finite-dimensional to avoid the technicalities
coming with unbounded operators. They give rise to domain problems and require a specific choice of
operator norm --- see \cite{SRF04b}. 

The present point of view differs from the usual one,
which starts from the Hilbert space generated by the density matrices. Instead, the GNS-representation is
used because it is more suited for later generalizations.
The main goal of the present work is precisely to present those results for which one would like
to find generalizations in the infinitely-dimensional case.

The manifold $\Mo$ of faithful quantum states can be parametrized in many ways.
It is tradition to label each quantum state $\omega$ by a corresponding density matrix $\rho$.
Here, the parameter-free approach of Pistone and coworkers \cite{PS95,PR99,PG13,MP17} is followed.
In particular, with each element $\omega_\rho$ of $\Mo$ is associated a chart centered at $\omega_\rho$.
Two atlasses are introduced. The atlas with the charts $\xi_\rho$, introduced in Section \ref {sect:atlas},
is technically less complicated. It turns $\Mo$ into a Banach manifold.
However, it is not linked in a straightforward manner with the Riemannian metric induced by
Bogoliubov's inner product. 
Therefore, another set of charts, denoted $\chi_\rho$, is introduced in Section \ref {sect:alt}.
A link between the charts $\xi_\rho$ and $\chi_\rho$ is found in Proposition \ref {exp:prop2}.

The dually affine connections are shortly mentioned in Sections \ref {sect:mix}, \ref {sect:exp}.
In the case of the exponential connection the charts $\chi_\rho$ provide affine coordinates.


\appendix

\section*{The GNS-representation of a matrix algebra}

This appendix is added for convenience of the reader.
Its content is well-known.

Choose an orthonormal basis of eigenvectors $\psi_n$, $n=1,2,\cdots,N$, of the 
strictly positive density matrix $\rho$.
It can be written as
\be
\rho=\sum_{n=1}^N p_n E_n
\quad\mbox{ where } \, E_n\, \mbox{ is the orthogonal projection onto }\, \Co\psi_n.
\nnee
Introduce now the vector $\Omega_\rho$ in ${\cal H}\otimes{\cal H}$ defined by
\be
\Omega_\rho=\sum_{n=1}^N\sqrt {p_n} \psi_n\otimes \psi_n.
\nnee
It satisfies $||\Omega_\rho||^2=\sum_{n=1}^Np_n=1$.
A short calculation shows that
\be
\Tr\rho A&=&((A\otimes \Io)\Omega_\rho,\Omega_\rho)
\quad\mbox{ for all } \, A\in{\cal A}.
\nnee
By assumption all eigenvalues $p_n$ are strictly positive.
Therefore $||A\Omega_\rho||=0$ implies $A=0$. This shows that $\Omega_\rho$ is separating.
Let $E_{n,m}$ be the orthogonal matrix which maps $\psi_m$ onto $\psi_n$.
It belongs to $\cal A$ and satisfies
\be
E_{n,m}\Omega_\rho&=&\sqrt{p_m}\psi_n\otimes\psi_m.
\nnee
This shows that ${\cal A}\Omega_\rho$ equals all of ${\cal H}\otimes{\cal H}$.
Hence, $\Omega_\rho$ is a cyclic vector for ${\cal A}\otimes\Io$. Because the GNS-representation is
unique up to unitary equivalence  one concludes that $A\mapsto A\otimes\Io$, together
with the Hilbert space ${\cal H}\otimes{\cal H}$ and the vector $\Omega_\rho$ is equivalent.

Note that the commutant  of ${\cal A}\otimes\Io$ equals $\Io\otimes{\cal A}$.

Introduce now an anti-linear operator $J$ defined by
\be
J(A\otimes\Io)\Omega_\rho=(\Io\otimes A')\Omega_\rho
\nnee
with $A'$ given by
\be
A'\psi_n&=&\sum_m(A^*\psi_m,\psi_n)\psi_m.
\nnee
From $||A'\psi_n||=||A\psi_n||$ then follows that $J$ is an isometry.
A short calculation shows that for all $A,B$
\be
(J(A\otimes\Io)\Omega_\rho,J(B\otimes\Io)\Omega_\rho)
&=&
((B\otimes\Io)\Omega_\rho,(A\otimes\Io)\Omega_\rho).
\nnee

Next define an anti-linear operator $S$ by $S(A\otimes \Io)\Omega_\rho=(A^*\otimes\Io)\Omega_\rho$.
This is the modular conjugation operator. One verifies immediately that the conjugate $F$ of $S$
satisfies $F(\Io\otimes A)\Omega_\rho=(\Io\otimes A^*)\Omega_\rho$ for all $A$.

The modular operator $\Delta$ by definition equals $S^*S=FF^*$.
Let us verify that $\Delta=\rho\otimes\rho^{-1}$.
It suffices to show that $\rho^{-1/2}\otimes\rho^{1/2}J=S$.
One calculates
\be
(\rho^{-1/2}\otimes\rho^{1/2})J(A\otimes \Io)\Omega_\rho
&=&
(\rho^{-1/2}\otimes\rho^{1/2})(\Io\otimes A')\Omega_\rho\cr
&=&
\sum_n\sqrt{p_n}(\rho^{-1/2}\otimes\rho^{1/2})\psi_n\otimes A'\psi_n\cr
&=&
\sum_n\psi_n\otimes \rho^{1/2}A'\psi_n\cr
&=&
\sum_{m,n}(A^*\psi_m,\psi_n)\psi_n\otimes \rho^{1/2}\psi_m\cr
&=&
\sum_{m,n}\sqrt{p_m}(A^*\psi_m,\psi_n)\psi_n\otimes\psi_m\cr
&=&
\sum_m\sqrt{p_m}A^*\psi_m\otimes\psi_m\cr
&=&(A^*\otimes\Io)\Omega_\rho\cr
&=&S(A\otimes\Io)\Omega_\rho.
\nnee

\end{document}